# Enhancement of Ce/Cr Codopant Solubility and Chemical Homogeneity in TiO$_2$ Nanoparticles through Sol-Gel *versus* Pechini Syntheses

Wen-Fan Chen[1*], Sajjad S. Mofarah[1], Dorian Amir Henry Hanaor[1,2], Pramod Koshy[1], Hsin-Kai Chen[1], Yue Jiang[1], and Charles Christopher Sorrell[1]

[1]School of Materials Science and Engineering, UNSW Sydney, Sydney, NSW    2052, Australia

[2]Fachgebiet Keramische Werkstoffe, Technische Universität Berlin, Berlin    10623, Germany

*Corresponding Author:    w.chen@unsw.edu.au

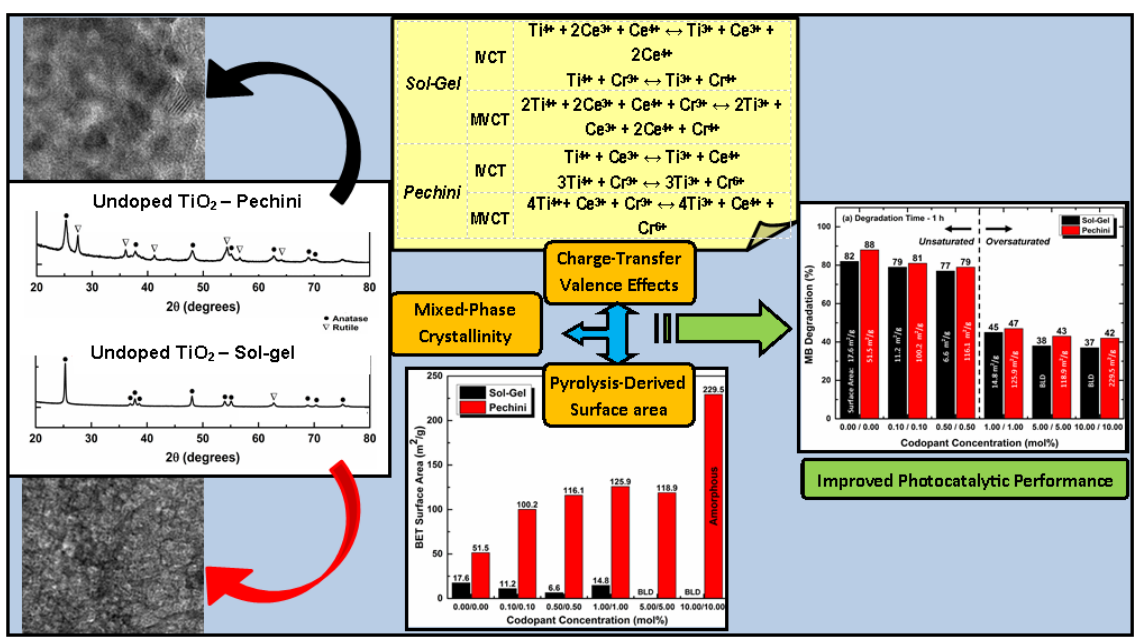

**Synopsis**

The Pechini method was more effective in fabricating Ce/Cr codoped TiO$_2$ nanoparticles of superior photocatalytic properties in comparison with the sol-gel method.  The improved efficiency was attributed to:   a) formation of mixed-phase anatase + rutile with greater homogeneity of dopant incorporation, b) development of open agglomerated morphologies with significantly higher surface areas, and c) favourable valence changes (increased Ti$^{3+}$ concentrations) owing to electron exchange in inter valence and/or multi valence charge transfer reactions.

**Abstract**

Ce/Cr codoped $TiO_2$ nanoparticles were synthesized using sol-gel and Pechini methods with heat treatment at 400°C for 4 h. A conventional sol-gel process produced well crystallized anatase while Pechini synthesis yielded less ordered mixed-phase anatase + rutile; this suggests that the latter method enhances Ce solubility, increases chemical homogeneity, but destabilizes the $TiO_2$ lattice. Greater structural disruption from the decomposition of the Pechini precursor formed more open agglomerated morphologies while the lower levels of structural disruption from pyrolysis of the dried sol-gel precursor resulted in denser agglomerates of lower surface areas. Codoping and associated destabilization of the lattice reduced the binding energies in both powders. $Cr^{4+}$ formation in sol-gel powders and $Cr^{6+}$ formation in Pechini powders suggest that these valence changes derive from synergistic electron exchange from intervalence and/or multivalence charge transfer. Since Ce is too large to allow either substitutional or interstitial solid solubility, the concept of *integrated solubility* is introduced, in which the Ti site and an adjacent interstice are occupied by the large Ce ion. The photocatalytic performance data show that codoping was detrimental owing to the effects of reduced crystallinity from lattice destabilization and surface area. Two regimes of mechanistic behavior are seen, which are attributed to the unsaturated solid solutions at lower codopant levels and supersaturated solid solutions at higher levels. The present work demonstrates that the Pechini method offers a processing technique that is superior to sol-gel because the former facilitates solid solubility and consequent chemical homogeneity. **Keywords:    $TiO_2$, Anatase, Rutile, Codoping, Sol-Gel, Pechini**

## 1.  Introduction

The use of sol-gel methods to obtain pure and doped oxide systems has become common practice because they offer precise compositional and microstructural control and versatility in homogeneity relative to more conventional solid-state syntheses. Sol-gel chemistry is an umbrella term encompassing a broad range of approaches. Most commonly, sol-gel syntheses of oxides involve the steps of hydrolysis and condensation of solutions derived from organometallic compounds and metal salts. However, there are some disadvantages to these processes, including the unavailability of suitable commercial precursors, moisture sensitivity, and differential hydrolysis kinetics, the latter two of which can result in inconsistent hydrolysis, chemical segregation, inhomogeneous precipitation, and agglomeration prior to heat treatment.[1] An alternative approach that can overcome these potential shortcomings involves polymer complexation by the Pechini route.[2,3] This methodology relies on the polyesterification of chelated cation species to produce homogenous polymeric precursors in which the precipitation of secondary phases is inhibited through steric entrapment, thereby maintaining molecular-level mixing at all stages of processing. This method generally employs carboxylate chelating agents and glycol groups, with the most common combination involving citrate-acid-driven chelation with ethylene-glycol-driven polyesterification.

Bonding in conventional sol-gel methods involves interconnected metal-oxygen M-O polymer networks formed by hydrolysis and condensation of metal alkoxides or salts.[4] In the formation of these metal-oxo networks, segregation of cationic species and formation of nanocrystals of binary compounds can occur owing to moisture sensitivity and/or preferential hydrolysis kinetics of individual metal precursors. In contrast, bonding in the Pechini method occurs through a polymer interconnected by covalent and coordinate bonding established during polyesterification between metal cations chelated with carboxylate groups and polyhydroxy alcohols.[5] This method traps cations in a polymer gel that retains a uniform cation distribution during mixing, calcination, pyrolysis, and crystallization. Figure 1 shows the schematic of Pechini process. Table 1 summarizes the differences between these two methods.

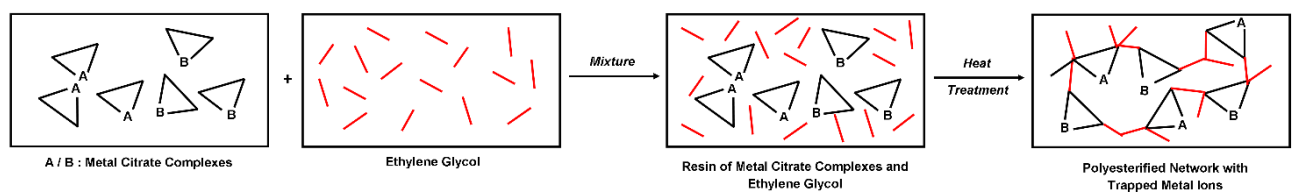

**Figure 1.** Schematic of the Pechini method[5]

**Table 1:** Summary of differences between solid-state, sol-gel, and Pechini methods

| Method | Process | Homogeneity | Particles |
|---|---|---|---|
| *Solid-State* | Salt Decomposition or Oxide Reaction | • Inhomogeneous Mechanical Mixture <br> • Variable Diffusivities – Chemical Gradients | • Agglomeration and Variable Grain Sizes <br> • Low Surface Areas |
| *Sol-Gel* | Hydrolysis & Condensation | • Homogeneous Sol Precursor <br> • Risk of Segregation and Precipitation | • Agglomeration and Uniform Large Grains <br> • Low Surface Areas |
| *Pechini* | Polyesterification of Chelated Cation Species | • Homogenous Polymeric Precursor <br> • Inhibition of Segregation and Precipitation | • Agglomeration & Uniform Small Grains <br> • High Surface Areas |

Table 2 provides a comprehensive summary of previous work reporting syntheses of doped $TiO_2$ by the Pechini method; there appear to be no reports of the synthesis of codoped $TiO_2$ powders by this method.

**Table 2**: Comprehensive summary of literature on the synthesis of doped TiO$_2$ powders by the Pechini method

| Dopant | Dopant Levels (mol%) | TiO$_2$ Polymorphs Observed | Crystallite Size (nm) | BET (m$^2$/g) | Pyrolysis Conditions | Recrystallization Conditions | Reference |
|---|---|---|---|---|---|---|---|
| Ag | 0-10 | Anatase | 9-11 | 54-83 | 400°C for 2 h | -- | 6 |
| Al |  |  | 10-12 | 70-100 |  |  |  |
| Ce |  |  | 6-12 | 58-108 |  |  |  |
| Nb |  |  | 10-11 | 63-104 |  |  |  |
| Al | 0-0.3 | Anatase | 3.5-9.7 | -- | 350°C for 30 h | 450°C for 15 h | 7 |
| Fe | 0-0.2 |  | 3.5-4.2 | -- |  | 550°C for 15 h |  |
| Fe | 0 | Anatase + Brookite | 20-27 | -- | 450°C for 2 h | -- | 8 |
|  | 0.3-3.0 | Anatase + Rutile | 5-88 |  |  |  |  |
| Zr | 0.25 | Anatase | 4.8-14.6 | 25-149 | 400°C for 1 h | 500°-900°C for 1 h | 9 |

A few studies into the effects of codoping with both transition and rare earth metals in TiO$_2$, have been reported using sol-gel methods.[10-13] Fe/Eu codoped anatase of crystallite size ~4.5-7.5 nm and surface area ~160-196 m$^2$/g showed that Fe (as hole trap) and Eu (as electron trap) acted synergistically to improve the photocatalytic performance through enhanced charge separation and interfacial charge transfer.[10] Although not stated, this mechanism involves intervalence charge transfer[11-13] since the proposed respective redox reactions are Fe$^{3+}$ → Fe$^{4+}$ and Eu$^{3+}$ → Eu$^{2+}$. Fe/Ho codoped anatase of crystallite size ~12.5-18.5 nm suggested that the Fe ions enhanced optical absorbance and the Ho ions retarded electron-hole recombination.[14] Fe/La codoped anatase-rutile mixtures of crystallite size ~10-30 nm and surface area ~25-84 m$^2$/g also acted synergistically to enhance radiation absorption and the corresponding photocatalytic performance.[15] V/La codoped anatase-rutile mixtures of crystallite size ~7-34 nm and surface area ~42-217 $m^2/g$ enhanced optical absorbance, retarded electron-hole recombination, and improved the photocatalytic performance.[16]

In the present work, Ce (rare earth metal) and Cr (transition metal) were used as dopants owing to their unfilled f and d shells, respectively, which have the potential to facilitate intervalence charge transfer between dopants and/or between matrix and dopant(s), which may improve photocatalytic performance.[11-13] Ce/Cr codoped nanoparticles were prepared at codopant concentrations up to 10.00 mol% each *via* (1) sol-gel and (2) Pechini methods, followed by thermal treatment at 400°C for 4 h. The mineralogical, morphological, surface, structural, chemical properties, optical, and photocatalytic characteristics were investigated in order to differentiate the effects of the synthesis method on these parameters.

## 2. Experimental Procedure

### 2.1 Synthesis of TiO$_2$ photocatalysts

Figure 2 shows the flow chart of preparation of TiO$_2$ powders through (1) sol-gel method and (2) Pechini method.

*(1) Sol-Gel Method*

A TiO$_2$ precursor solution was prepared by dissolving titanium tetra-isopropoxide (TTIP, Reagent Grade, 97 wt%) in isopropanol (Reagent Plus, 99 wt%) to obtain a 0.1 M solution (2.8 g of TTIP were diluted with isopropanol to 100 mL volume). Quantities of cerium (III) nitrate hexahydrate Ce(NO$_3$)$_3$·6H$_2$O (Sigma-Aldrich, ≥99 wt%) and chromium (III) nitrate nonahydrate Cr(NO$_3$)$_3$·9H$_2$O (Sigma-Aldrich, ≥99 wt%) corresponding to each of the selected dopant levels (metal basis) were added to each solution. All reagents were purchased from Sigma-Aldrich (Sydney, Australia). Precursor solutions were stirred magnetically at 200°C for 20 h to effect hydrolysis, condensation, and drying. The resultant colloidal powders were transferred to alumina crucibles and heat treated at 400°C for 4 h (The heating rates were 1°C/min from room temperature to 400°C, followed by natural cooling over ~10 h). The heat treated powders then were ground to nanoparticles by tungsten carbide ring milling (Rocklabs, Auckland, New Zealand) for 1 min.

*(2) Pechini Method*

Solution 1: Aqueous Ti citrate solutions were prepared by combining appropriate amounts of titanium tetra-isopropoxide (TTIP), citric acid (CA), and deionized (DI) water, followed by magnetic stirring at 65°C for 24 h to obtain a 0.1 M citrate-chelated solution. Solutions 2 and 3: Aqueous Ce and Cr citrate solutions were prepared by combining quantities of Ce(NO$_3$)$_3$·6H$_2$O or Cr(NO$_3$)$_3$·9H$_2$O with CA and DI water to obtain a 0.005 M citrate-chelated solution for each precursor. For all three solutions, the molar composition Ti:CA = 1:3 allowed full chelation of the Ti$^{4+}$ cations by the citrate anions. Quantities of Solutions 2 and 3 were added to Solution 1 and mixed by magnetic stirring at room temperature for 5 min, after which polyesterification (to form solid polymeric resin) was effected by adding ethylene glycol (EG) at the molar ratio Ti:CA:EG = 1:3:9, followed by magnetic stirring at 200°C for 20 h. This molar ratio was found to be sufficient to result in adequate dispersion of the cations in the resultant polymer network. As the water evaporated, a resin containing sterically immobilised metal cations was formed without precipitation. The resultant resins were transferred to alumina crucibles and heat treated at 400°C for 4 h (The heating rates were 1°C/min from room temperature to 400°C, followed by natural cooling over ~10 h). The heat treated powders then were ground to nanoparticles by tungsten carbide ring milling.

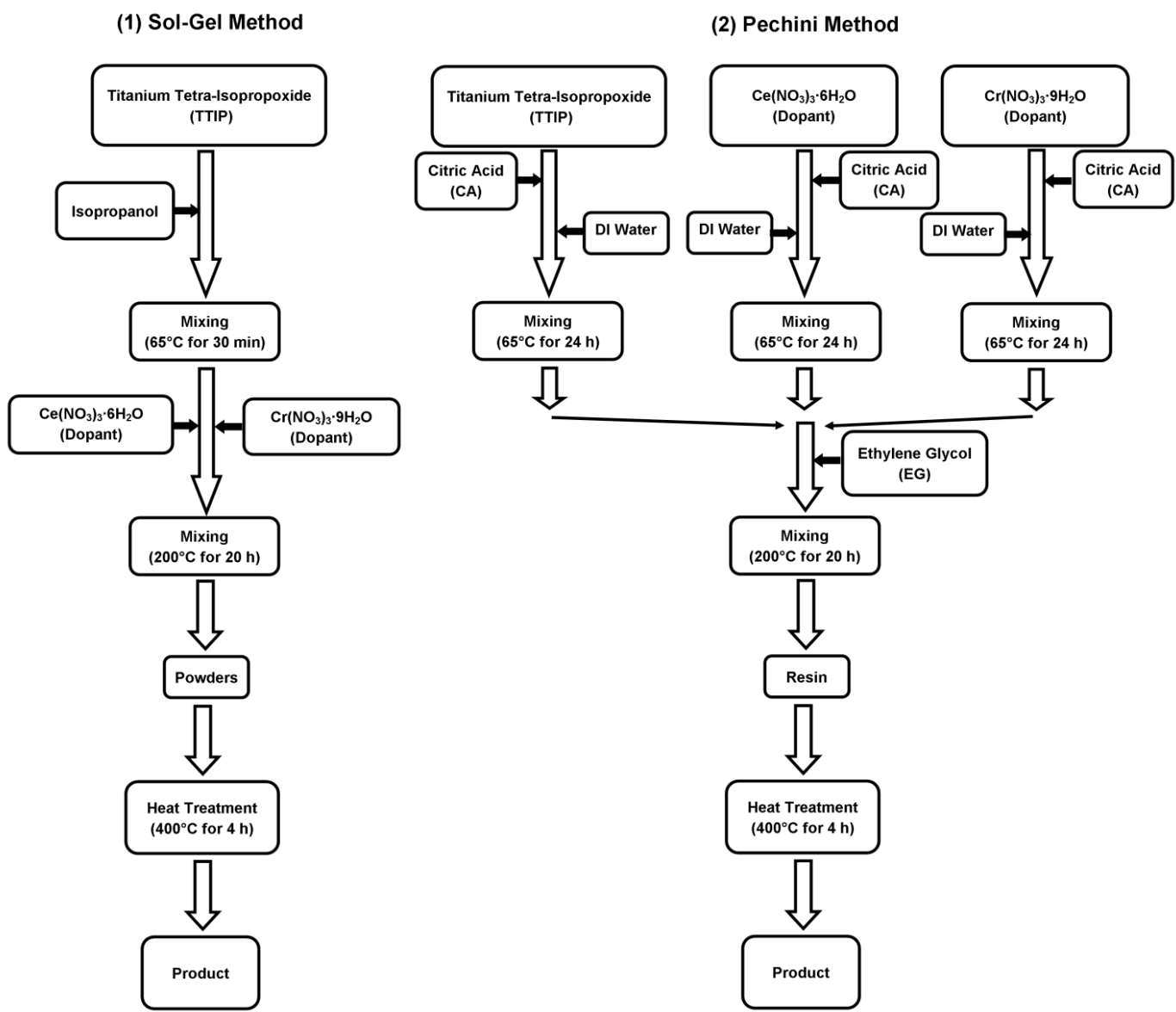

Figure 2. Flow chart for codoped TiO$_2$ powders fabricated by sol-gel and Pechini methods

## 2.2 Characterization

The resultant nanoparticles were characterized using the following techniques shown in Table 3. The photocatalytic activity was assessed by photobleaching of methylene blue (MB, M9140, dye content ≥82 wt%, Sigma-Aldrich) solutions of $10^{-5}$ M concentration. Suspensions of the sol-gel powders, Pechini powders, and Degussa P25 powder were prepared by adding 8 mg of each powder and 8 mL of MB solution to a 25 mL Pyrex beaker. Each suspension was soaked for 15 min in a light-obstructing container, after it was exposed to UV radiation (UVP, 3UV-38, 8 W, ~10 cm light-liquid distance, 365 nm wavelength) with constant magnetic stirring (300 rpm) for 1 h. The solutions then were isolated from the powders by centrifugation for 10 min (5000 rpm), after which the they were analyzed by ultraviolet-visible absorbance spectrophotometry (UV-Vis, PerkinElmer Lambda 35 UV-Visible Spectrometer, aperture 20 mm × 10 mm) at 664 nm, which is the maximal absorption wavelength of MB.

**Table 3:** Details of analytical instrumentation

| Analytical Instruments | Acronym | Parameters |
| --- | --- | --- |
| *X-Ray Diffraction* | XRD | PANalytical Xpert Multipurpose X-ray Diffractometer, CuKα, step size 0.026° 2θ, step speed 29.07° 2θ/min |
| *Scanning Electron Microscopy* | SEM | FEI Nova SEM 230/450, 5 kV |
| *Brunauer-Emmett-Teller Analysis* | BET | TriStar II Plus 2.02 |
| *Field Emission Transmission Electron Microscopy* | FETEM | Philips CM200, 200 kV |
| *X-Ray Photoelectron Spectroscopy* | XPS | Thermo Scientific ESCALAB 250Xi, 500 μm beam diameter |
| *Fourier Transform Infrared Spectrometry* | FTIR | PerkinElmer Spotlight 400 |

## 3. Results and Discussion

Figure 3 shows XRD patterns of codoped TiO$_2$ powders following heat treatment at 400°C for 4 h. While the sol-gel powders crystallized solely as anatase, which is consistent with other reports,[11,12,17] the Pechini powders consisted of mixed anatase and rutile, which, as shown in Table 2, is the case only for Fe-doped TiO$_2$. It is notable that 0-10 mol% Ce-doped TiO$_2$ annealed at 400°C for 2 h consisted of anatase only.[6] No secondary oxide phases were detected. The anatase peak intensities of both powders appeared to decrease with increasing dopant levels. However, the rutile peak intensity of the Pechini powder upon initial doping decreased but increasing dopant levels appeared to increase slightly the peak intensities. Consequently, Ce solubility destabilized both anatase and rutile and does not serve as a source of nucleation sites. Since this effect was more significant in the Pechini powders, this suggests that the Pechini method provides better chemical homogeneity and hence a superior mechanism of solubility. This conclusion is supported by the observation of rutile in the Pechini powders annealed at a relatively low temperature, which is indicative powders of greater reactivity.

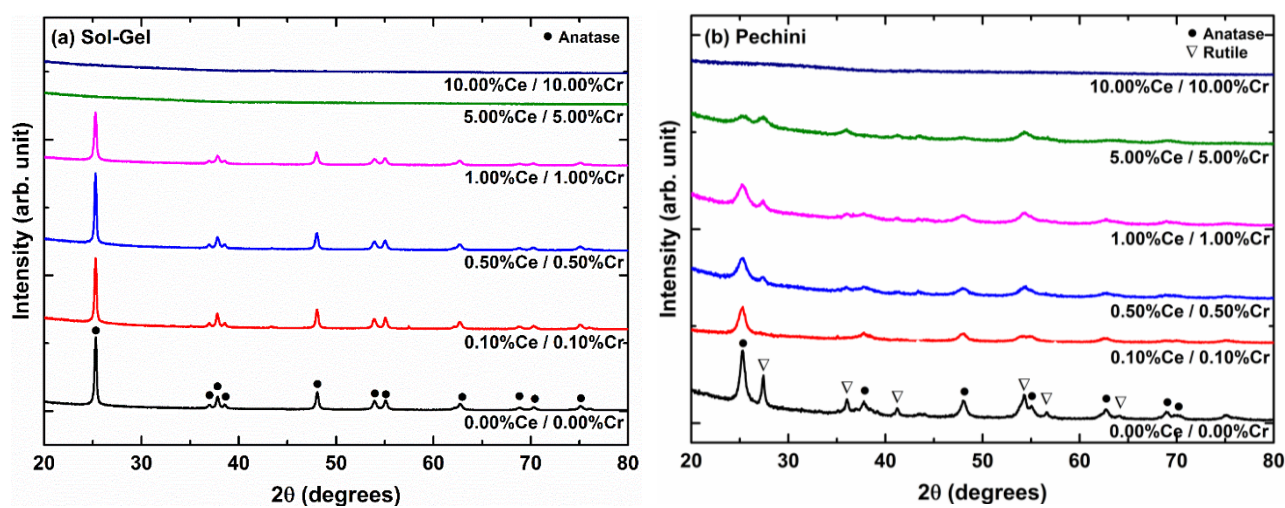

**Figure 3.** XRD patterns of codoped TiO$_2$ powders fabricated by heat treatment at 400°C for 4 h: (a) sol-gel, (b) Pechini

The Pechini powders also suggest that Cr and/or Ce may have promoted the anatase → rutile transformation although both cations are known to be inhibiters of rutile formation.[18] This contradiction can be explained by the synergistic effects of the two dopants by intervalence charge transfer (IVCT) and/or multivalence charge transfer (MVCT), as discussed subsequently.

The presence of rutile at the low heat treatment temperature of 400°C may have resulted from the heat generated by the pyrolysis of the polyester resin is possible, which would have facilitated the anatase → rutile transformation. Further, pyrolysis may have resulted in partial Ti$^{4+}$ → Ti$^{3+}$ reduction and the formation of associated oxygen vacancies, which also would have facilitated the mobility necessary for the reconstructive anatase → rutile transformation.[19]

Table 4 shows details of the principal anatase (101) and rutile (110) peaks for both methods of synthesis. Although no internal standard was used, the peak positions appeared not to shift with increasing codopant concentration for the sol-gel powders while peaks were shifted to significantly higher angles for the Pechini powders at codopant levels between 0.50% Ce / 0.50% Cr and 1.00% Ce / 1.00% Cr. Again, these data support the conclusion that the Pechini method provides a superior mechanism of solubility. Further support for this conclusion is present in the irregular and regular variations in crystallite size for the sol-gel and Pechini powders, respectively.

**Table 4.** Details of codoped $TiO_2$ powders fabricated by sol-gel and Pechini methods

|  | Codopant Level | | XRD Peak Position[A] | | Volume Fraction[B] | | Crystallite Size (nm)[B] | |
|---|---|---|---|---|---|---|---|---|
|  | *Ce (mol%)* | *Cr (mol%)* | *(101)$_{Anatase}$* (°2θ)* | *(110)$_{Rutile}$* (°2θ)* | *V$_{Anatase}$* (Vol%) | *V$_{Rutile}$* (Vol%) | *Anatase* | *Rutile* |
| Sol-Gel | 0 | 0 | 25.27 | 0 | 100 | 0 | 49.0 | -- |
|  | 0.10 | 0.10 | 25.29 | 0 | 100 | 0 | 53.5 | -- |
|  | 0.50 | 0.50 | 25.28 | 0 | 100 | 0 | 53.5 | -- |
|  | 1.00 | 1.00 | 25.28 | 0 | 100 | 0 | 39.0 | -- |
|  | 5.00 | 5.00 | 0 | 0 | 0 | 0 | -- | -- |
|  | 10.00 | 10.00 | 0 | 0 | 0 | 0 | -- | -- |
| Pechini | 0 | 0 | 25.26 | 27.39 | 69 | 31 | 14.5 | 30.0 |
|  | 0.10 | 0.10 | 25.25 | 27.39 | 97 | 3 | 9.5 | (2.0) |
|  | 0.50 | 0.50 | 25.25 | 27.30 | 82 | 18 | 8.5 | 18.0 |
|  | 1.00 | 1.00 | 25.33 | 27.36 | 78 | 22 | 8.0 | 15.5 |
|  | 5.00 | 5.00 | 25.45 | 27.40 | 49 | 51 | 7.5 | 8.5 |
|  | 10.00 | 10.00 | 0 | -- | 0 | -- | -- | -- |

[A] No internal standard used

[B] Calculated by the Spurr and Myers method[20]

[C] Calculated by the Scherrer method[21]

It is notable that the XRD peak positions for the Pechini powders indicated lattice contraction, which would appear to be unlikely owing to the large size of the lanthanide Ce. The relevant sixfold Shannon crystal radii[22] are given in Table 5. According to Hume-Rothery's rules[23, 24], substantial solid solubility requires the dopant ion to be of radius ≤15% different from that of the matrix ion. Accordingly, Ce should not be soluble and only $Cr^{3+}$ and $Cr^{4+}$ should ($Cr^{2+}$ is not considered owing to its thermodynamic instability[13]). If Ce exhibits interstitial solubility, then it would have to be accommodated in one or both of the interstices located above and below the central Ti ion in the elongated $TiO_6$ octahedron[11]. Since Table 4 shows that this interstice has a radius of 0.0782 nm, it is too small to accommodate a Ce ion. However, if solubility integrates both substitutional and interstitial mechanisms to occupy both types of sites, then the lengths available would be 0.0782 + 0.0745 = 0.1527 nm (interstice + Ti site) or 0.0782 + 0.0745 + 0.0782 nm = 0.2309 nm (two interstices + Ti site). Both of these are sufficiently large to accommodate a Ce ion. This *integrated solubility* would be likely to result in expansion of the *a-b* plane and contraction of the *c* axis. Assuming a hard sphere model, integrated interstice + Ti site, and $Ce^{4+}$ solubility, the lattice distortion would be expansion of the *a* or *b* axis of 0.0375 nm and a contraction of the *c* axis of 0.1034 nm. These represent unit cell[18] alterations of ~10% *a* or *b* axis expansion and ~11% *c* axis contraction. Thus, the XRD data suggest that *c* axis contraction provides the dominant distortional effect, which has been suggested previously[11].

Table 5. Relevant sixfold Shannon crystal radii[22]

| Cation | Valence | Spin | Crystal Radius (nm) | Radius Difference with $Ti^{4+}$ (%) | Radius Difference with Interstice (%) |
|---|---|---|---|---|---|
| Ti | 4+ | -- | 0.0745 | -- | − 4.73 |
| Ti | 3+ | -- | 0.081 | -- | + 3.58 |
| Ti | 2+ | -- | 0.100 | -- | + 27.90 |
| Ce | 4+ | -- | 0.101 | + 35.57 | + 29.16 |
| Ce | 3+ | -- | 0.115 | + 54.36 | + 47.06 |
| Cr | 6+ | -- | 0.058 | − 22.15 | − 25.83 |
| Cr | 5+ | -- | 0.063 | − 15.44 | − 19.44 |
| Cr | 4+ | -- | 0.069 | − 7.38 | − 11.76 |
| Cr | 3+ | -- | 0.0755 | + 1.34 | − 3.45 |
|  | 2+ | Low | 0.087 | + 16.78 | + 11.25 |
|  |  | High | 0.094 | + 26.17 | + 20.20 |
|  | Interstice[11] |  | 0.0782 | + 4.97 | -- |

Figures 4 and 5 show SEM images of codoped TiO2 powders comprising the sol-gel and Pechini powders, respectively. The sol-gel powders were comprised largely of a consistent distribution of irregularly shaped and dense agglomerates, which exhibited a limited distribution density of small particles decorating the surfaces. Figure 4(d) shows a large amorphous particle in the background while Figures 4(e) and (f) show these more clearly. Figure 3(a) indicates that Figures 4(e) and (f) should consist solely of amorphous grains but this does not appear to be the case in Figure 4(e) owing to the presence of agglomerates. Reasons for this contradiction are that the amount of crystalline TiO$_2$ could be below the level of detection (BLD), which seems unlikely, or that the agglomerates are destabilized relicts of the crystalline agglomerates. The Pechini powders are somewhat different in that the agglomerates are rounded and they are comprised of much smaller particles. There also is more porosity between the agglomerates than is the case for the sol-gel powders. While Figure 5(f) shows a large amorphous particle, there also is a substantial amount of agglomerates. This is contrary to the XRD data shown in Figure 3(b), which indicates that the powder of amorphous. Again, these agglomerates are assumed to be destabilized by the dissolution of a dopant level beyond that which the structure can accommodate.

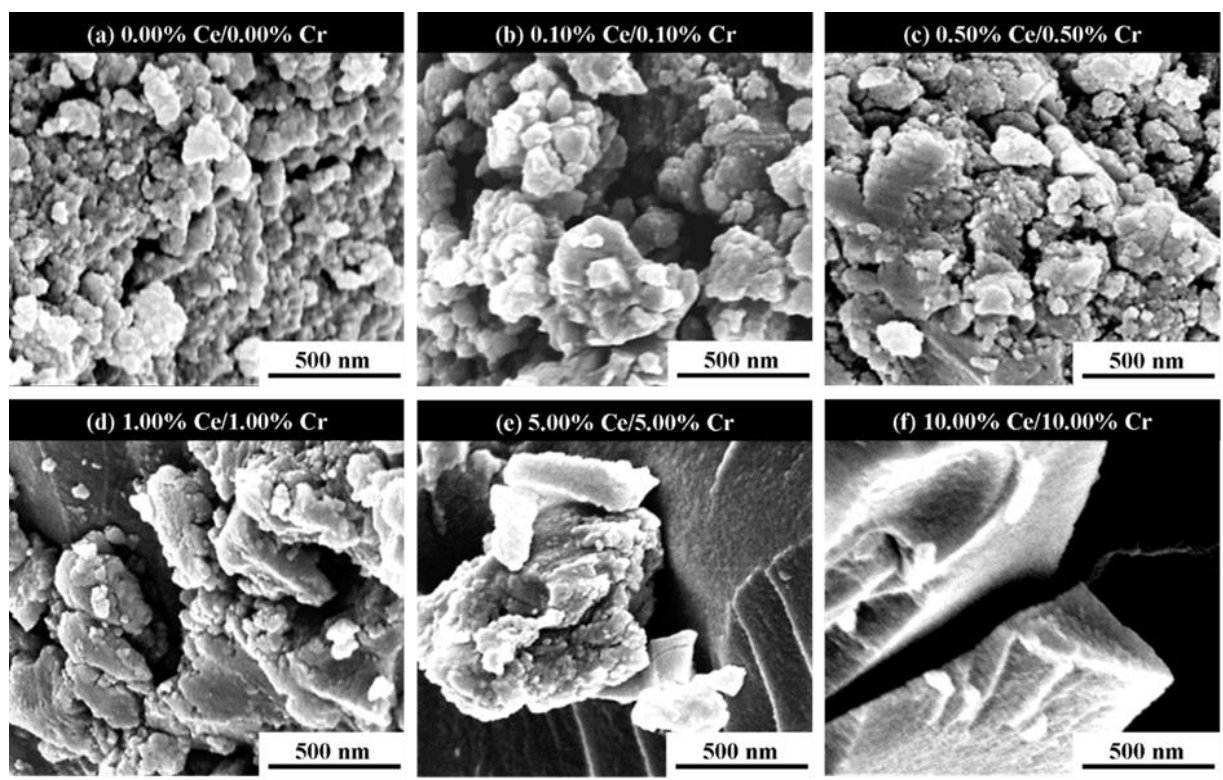

**Figure 4.** SEM images of codoped TiO$_2$ powders synthesized by sol-gel

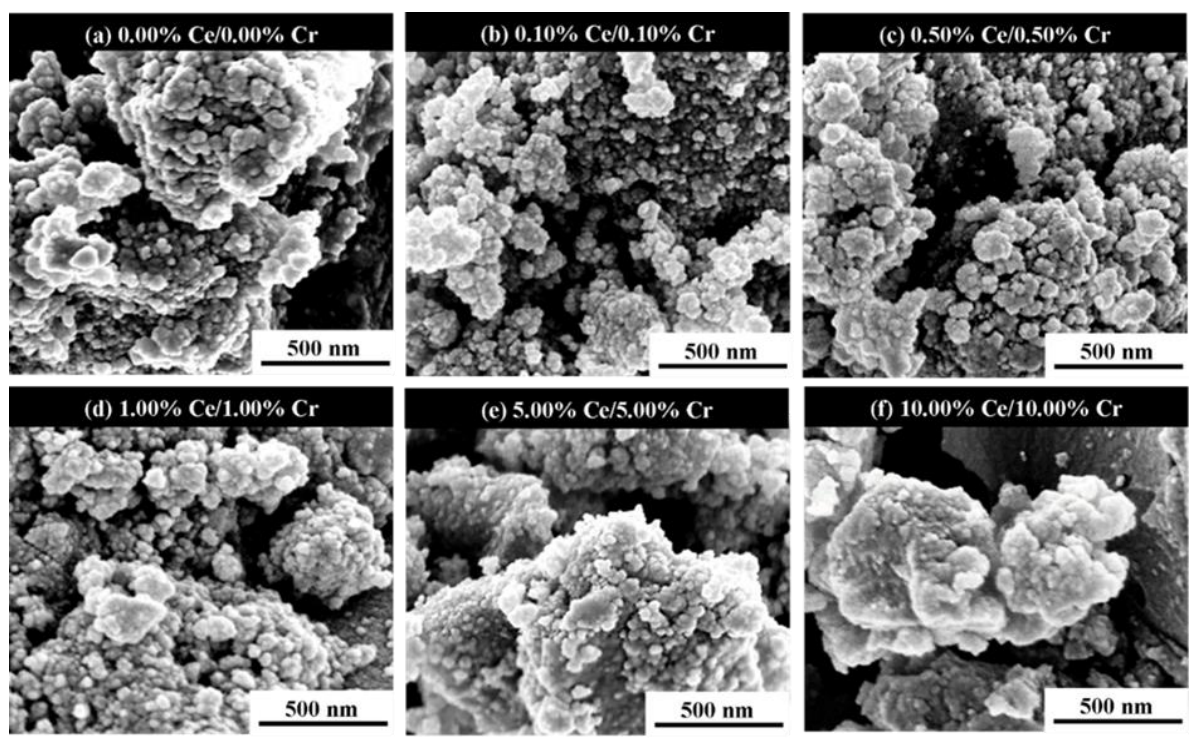

**Figure 5.** SEM images of codoped TiO$_2$ powders synthesized by the Pechini method

Figure 6 shows the BET surface areas of codoped $TiO_2$ powders comprising the sol-gel and Pechini powders. The sol-gel powders were of very low surface areas, which appeared to decrease with increasing codopant levels. At the point of structural collapse and resultant amorphization at ~5.00% Ce / 5.00% Cr, the surface area decreased to BLD. In contrast, the Pechini powders exhibited increasing surface areas up to ~0.50% Ce / 0.50% Cr, at which point they stabilized until structural collapse resulted in a significant increase. The last point supports the view that the major component is the agglomerates and that these are amorphous relicts.

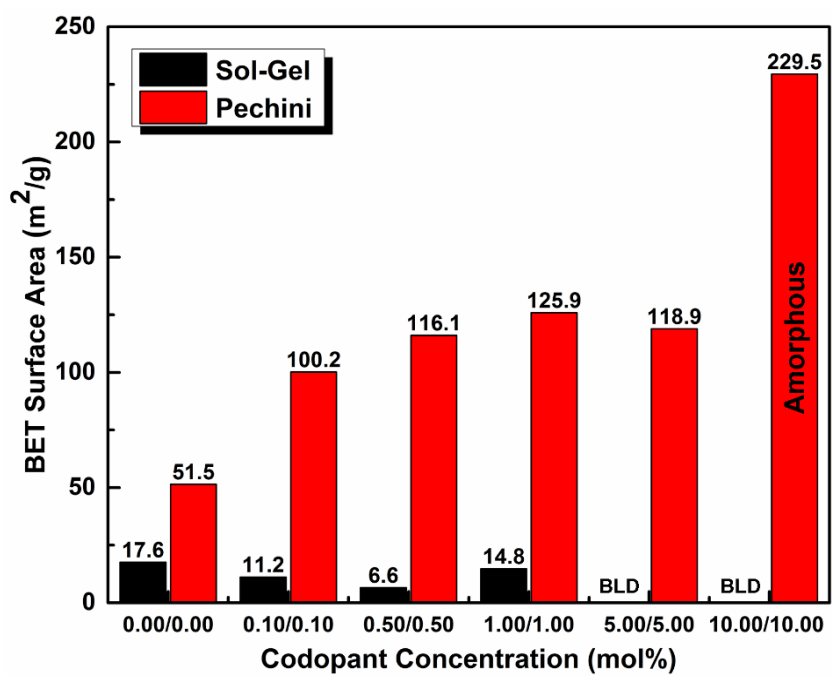

**Figure 6.  BET surface areas of codoped TiO$_2$ powders synthesized by sol-gel and Pechini methods**

The microstructures of the Pechini powders in Figure 5 and the surface areas in Figure 6 support the view that both of these were generated by disruption of the microstructure by gas exudation during decomposition of the polyester resin. Calculations on the basis of the raw materials reveal the following weight losses:

Sol-gel:    59.1 wt% carbon        0.3 wt% water

Pechini:    3.1 wt% carbon         89.4 wt% water

These data suggest that the greater structural disruption of the Pechini precursor derives not from the pyrolysis of the carbon components but from dehydroxylation.

FETEM and selected area electron diffraction (SAED) patterns for the codoped TiO$_2$ nanoparticles (undoped, 0.10% Ce / 0.10% Cr, 0.50% Ce / 0.50% Cr) comprising the sol-gel and Pechini powders are shown in Figures 7 and 8, respectively. The low-magnification images reveal that the Pechini powders exhibited greater surface areas both on the agglomerate surfaces and within the agglomerates, which confirms the comments for the SEM images. The high-magnification images show that there was no apparent expansion of the anatase lattice with increasing codopant levels while lattice expansion occurred in the Pechini powders, which confirms the data in Table 4. Also, lattice fringes from the (101) plane of anatase (yellow font) and the (110) plane of rutile (blue font) could be identified, revealing exposed facets consistent with those detected by XRD. The diffraction spots of the SAED patterns for the sol-gel powders confirm the presence of well crystallized anatase while the diffuse rings for the Pechini powders confirm their reduced crystallinity, in confirmation of the XRD data in Figures 3(a) and (b).

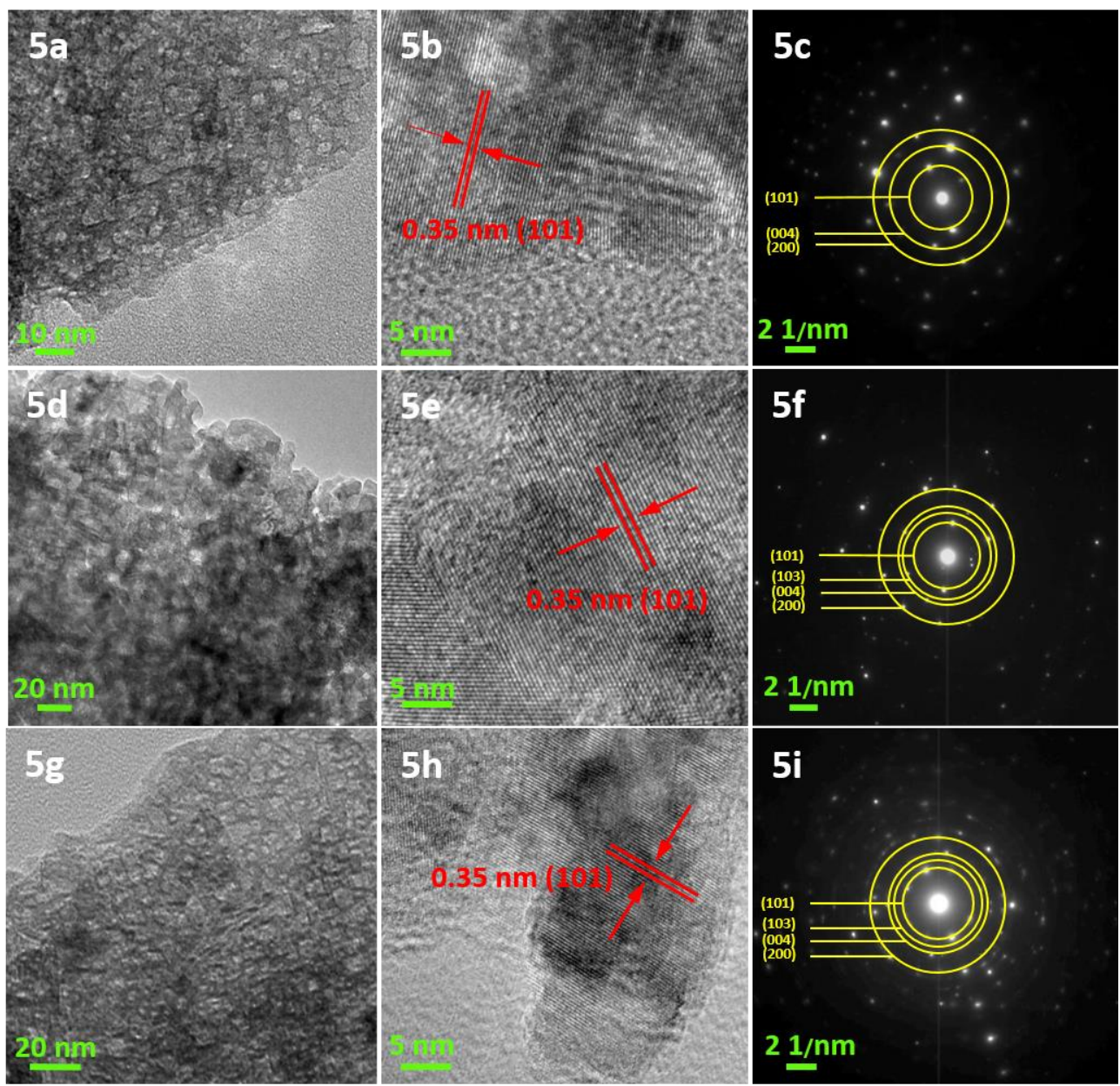

**Figure 7**. FETEM and SAED images of codoped TiO$_2$ powders synthesized by sol-gel method: (a-c) undoped, (d-f) 0.10 mol% Ce/0.10 mol% Cr, (g-i) 0.50 mol% Ce/0.50 mol% Cr

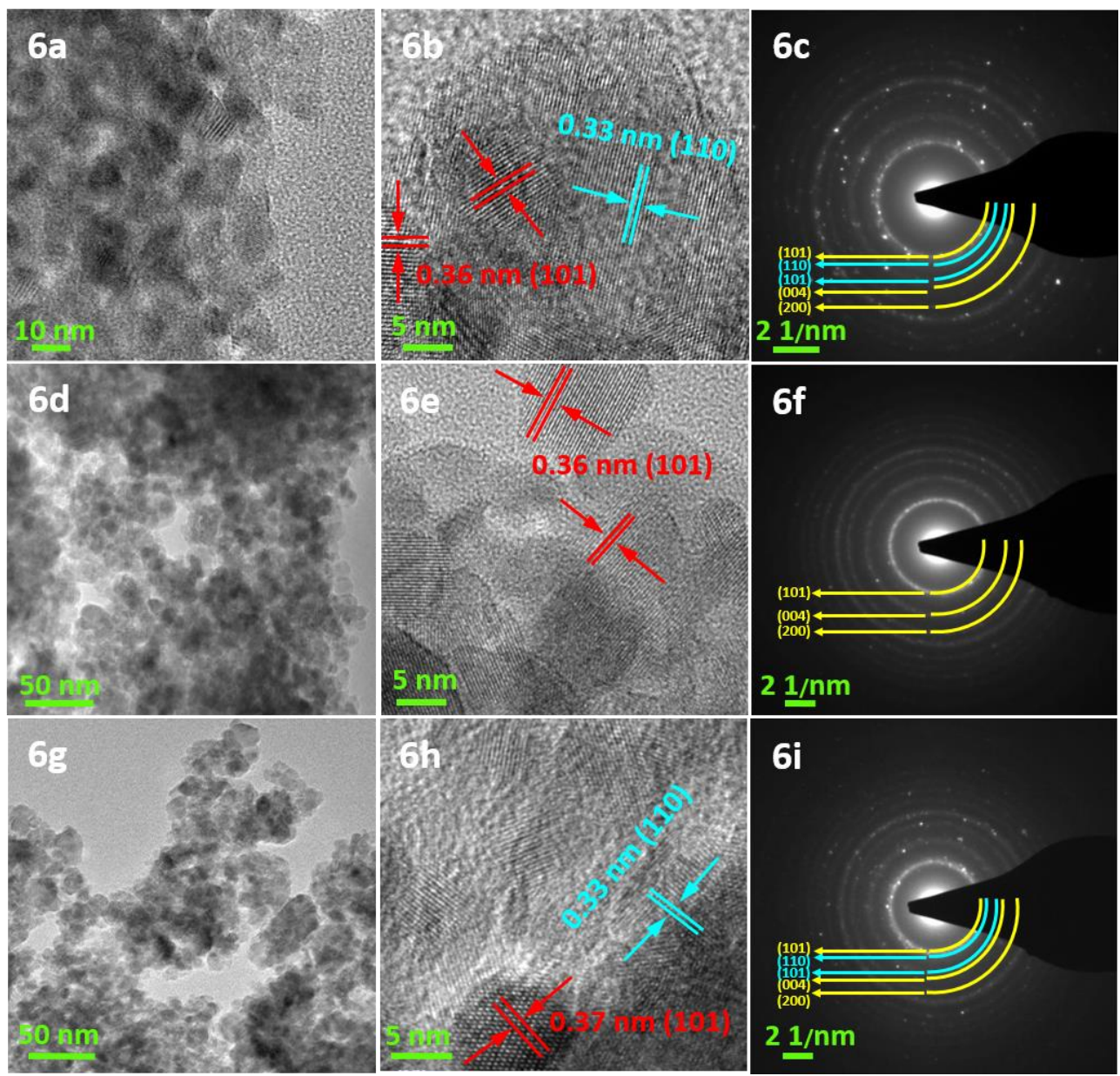

**Figure 8.** FETEM and SAED images of codoped anatase (yellow) and rutile (blue) TiO$_2$ powders synthesized by Pechini method: (a-c) undoped, (d-f) 0.10 mol% Ce/0.10 mol% Cr, (g-i) 0.50 mol% Ce/0.50 mol% Cr

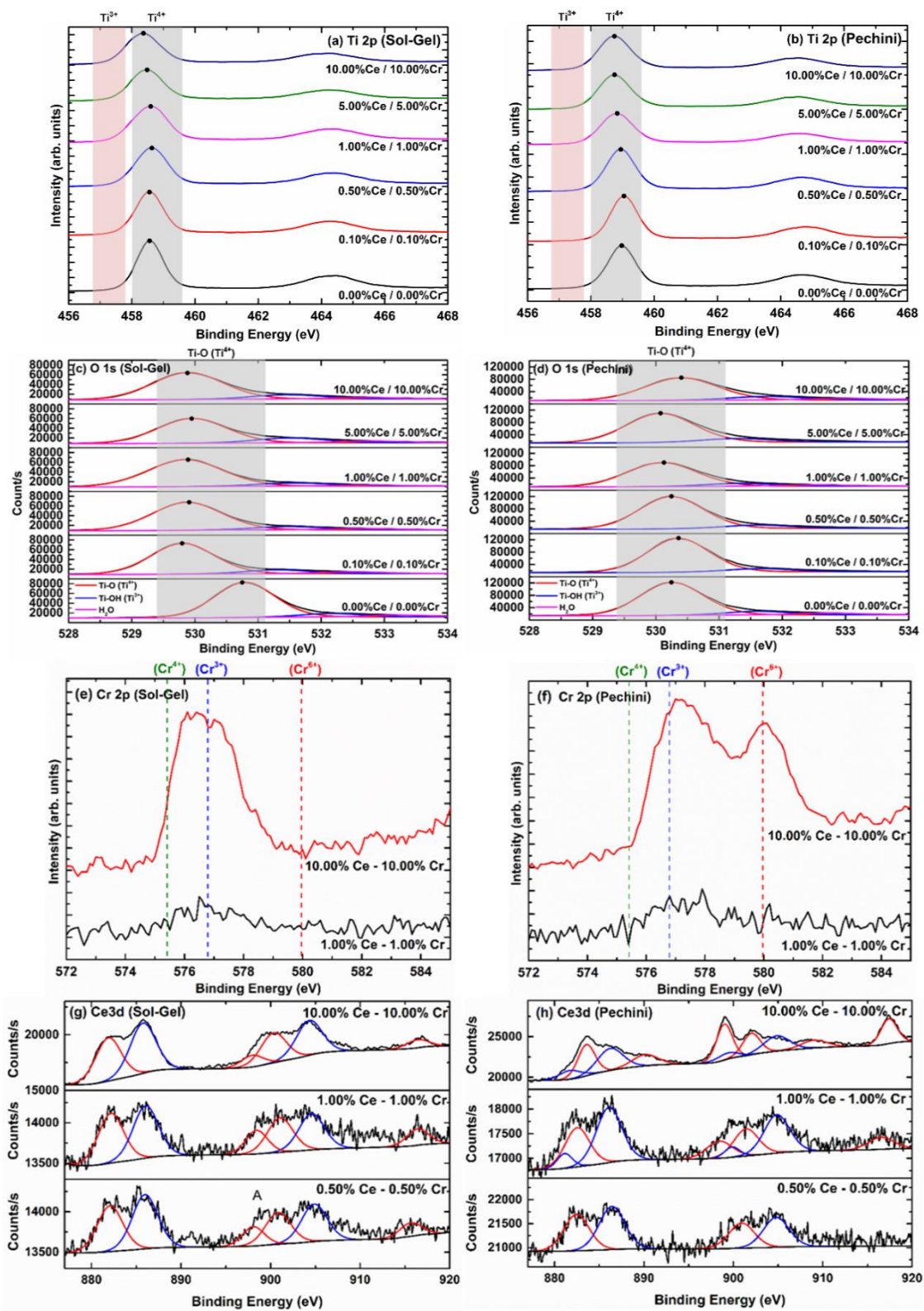

**Figure 9.** XPS spectra for TiO$_2$ powders synthesized by sol-gel and Pechini methods: (a, b) Ti2p, (c, d) O1s, (e, f) Cr2p, (g, h) Ce3d (Ce$^{3+}$ in blue, Ce$^{4+}$ in red)

Figure 9 shows the XPS spectra of codoped TiO$_2$ powders comprising the sol-gel and Pechini powders. These measurements were calibrated using the C 1s (284.5 eV) spectrum of adventitious carbon, which is present in all samples. Comparison of Figures 9(a) and (b) reveal that the undoped sol-gel powder had a lower binding energy than that of the Pechini powder, which typically would be considered to be indicative of the former's greater Ti$^{3+}$ (~457.0 eV[25]) content relative to that of Ti$^{4+}$ (~458.6 eV[25]). This is attributed to the reduction effect from pyrolysis of the sol-gel precursor, as discussed subsequently. Increasing codopant level was consistent for both powders, for which the XRD data in Figures 3(a) and (b) revealed that codoping destabilized the lattice, resulting in gradual reduction of the binding energy. where the Ti$^{3+}$ content increased and/or the lattice was destabilized. The peak shifts were slightly larger in the Pechini powders, again supporting the conclusion that the mechanism of solubility of the Pechini method is superior to that of the sol-gel method.

The O 1s data in Figures 9 (c) and (d) should allow differentiation between the Ti$^{3+}$ (~531.3 eV[25]) and Ti$^{4+}$ (~530.0 eV[25]) valence states. The data for the sol-gel powders in Figure 9 (c) suggest that Ce doping initially decreased the binding energy, after which increased dopant levels caused a smaller reverse trend. In contrast, the data for the Pechini powders in Figure 9 (d) show completely converse trends. These data are difficult to rationalize and so mitigate the previous comments concerning the Ti 2p XPS data. The probable reason for the uncertainty is the three variables, which are the Ce valences, the Cr valences, and the potential for retention of structural carbon, in addition to Ti$^{3+}$ content and lattice stability.

Figures 9 (e) and (f) indicate that the sol-gel powders contained Cr$^{3+}$ (~576.6 eV[25]) and Cr$^{4+}$ (~575.4 eV[25]) while the Pechini powders contained Cr$^{3+}$ and Cr$^{6+}$ (~580 eV[25]). Figures 9 (g) and (h) indicate that the Ce$^{3+}$ and Ce$^{4+}$ contents are not consistent. The FTIR spectra in Figure 10 show that both powders contain carbon-bonded groups but that the contents of these in the sol-gel powders are significantly lower than those in the Pechini powders. This is consistent with a lower extent of pyrolysis of the latter, as mentioned in conjunction with the data for the microstructures in Figure 5 and the surface areas in Figure 6. Consequently, the combination of five variables effectively precludes detailed interpretation of the O 1s XPS data.

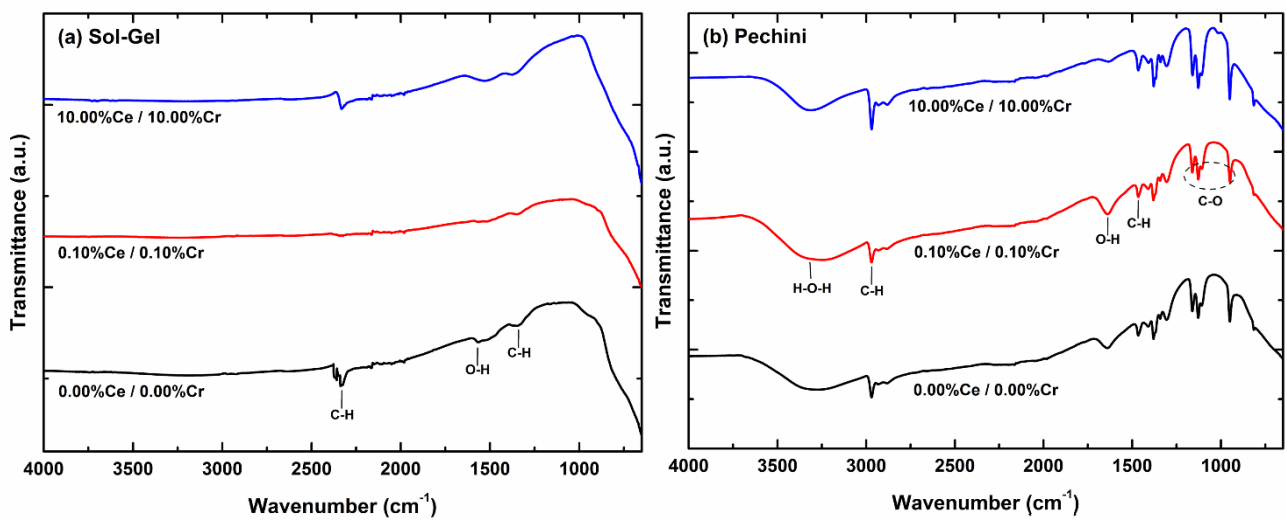

**Figure 10**. FTIR spectra of codoped TiO$_2$ powders synthesized by heat treatment at 400°C for 4 h: (a) sol-gel and (b) Pechini method (peak identifications: C-H, O-H[26], C-O[27])

The discussion of the XRD data focuses on intervalence charge transfer (IVCT). This concept is important because it explains why the valences of the starting materials can be altered, why reduced valence states are present following processing under oxidizing conditions, and why more thermodynamically stable valences can be altered to less stable valences. Table 6 summarizes the four potential IVCT mechanisms which that may occur during the codoping used in the present work.

**Table 6**: Potential IVCT mechanisms from doping and codoping of TiO$_2$

| IVCT Mechanisms | Potential IVCT Reactions |
|---|---|
| *Matrix Only* | $Ti^{4+} + Ti^{4+} \leftrightarrow 2Ti^{3+} + V_o^{\bullet\bullet}$ |
| *Matrix and Dopant* | $Ti^{4+} + Ce^{3+} \leftrightarrow Ti^{3+} + Ce^{4+}$ |
| | $Ti^{4+} + Cr^{3+} \leftrightarrow Ti^{3+} + Cr^{4+}$ |
| | $Ti^{4+} + Cr^{4+} \leftrightarrow Ti^{3+} + Cr^{5+}$ |
| | $Ti^{4+} + Cr^{5+} \leftrightarrow Ti^{3+} + Cr^{6+}$ |
| *Dopant Only* | $Ce^{3+} + Ce^{4+} \leftrightarrow Ce^{4+} + Ce^{3+}$ |
| | $Cr^{3+} + Cr^{4+} \leftrightarrow Cr^{4+} + Cr^{3+}$ |
| | $Cr^{3+} + Cr^{5+} \leftrightarrow Cr^{4+} + Cr^{4+}$ |
| | $Cr^{3+} + Cr^{6+} \leftrightarrow Cr^{4+} + Cr^{5+}$ |
|  | $Cr^{4+} + Cr^{6+} \leftrightarrow Cr^{5+} + Cr^{5+}$ |
|  | $Cr^{5+} + Cr^{6+} \leftrightarrow Cr^{6+} + Cr^{5}$ |
| ***Between Dopants*** | $Ce^{3+} + Cr^{4+} \leftrightarrow Ce^{4+} + Cr^{3+}$ |
|  | $Ce^{3+} + Cr^{4+} \leftrightarrow Ce^{4+} + Cr^{3+}$ |
|  | $Ce^{3+} + Cr^{5+} \leftrightarrow Ce^{4+} + Cr^{4+}$ |
|  | $Ce^{3+} + Cr^{6+} \leftrightarrow Ce^{4+} + Cr^{5+}$ |

\* $V_o^{\bullet\bullet}$ is an oxygen vacancy

However, since there are simultaneously variable valences for the matrix and two dopants, then multivalence charge transfer (MVCT)[28], which involves matrix and multiple dopants, also must be considered. Since the number of permutations for the electron exchanges between these three cations is large, only a simplified expression for the principle is given:

$$(Ti^{3+} \leftrightarrow Ti^{4+}) \leftrightarrow (Ce^{3+} \leftrightarrow Ce^{4+}) \leftrightarrow (Cr^{3+} \leftrightarrow Cr^{4+} \leftrightarrow Cr^{5+} \leftrightarrow Cr^{6+}) \qquad (1)$$

Table 7 shows the IVCT and MVCT mechanisms suggested by the XPS data. In the sol-gel powders, since $Ti^{3+/4+}$, $Ce^{3+/4+}$, $Cr^{3+/4+}$ were present in the codoped samples, then charge balance could be achieved only through the presence of multiple numbers of individual ions rather than just single numbers. If this is the case, then electron exchange occurs over multiple lattice sites and so is, as such, localized over more than two point defects. In the Pechini powders, since $Ti^{3+/4+}$, $Ce^{3+/4+}$, $Cr^{3+/6+}$ were present in the codoped samples, then there are two possibilities for charge balance through IVCT/MVCT: (1) $Cr^{4+}$ and/or $Cr^{5+}$ must have existed transiently or permanently below the level of detection (BLD) and/or (2) multiple numbers of individual ions were involved.

**Table 7**: IVCT and MVCT mechanisms suggested by XPS data

| ***Sol-Gel*** | IVCT | $Ti^{4+} + 2Ce^{3+} + Ce^{4+} \leftrightarrow Ti^{3+} + Ce^{3+} + 2Ce^{4+}$ |
|---|---|---|
|  |  | $Ti^{4+} + Cr^{3+} \leftrightarrow Ti^{3+} + Cr^{4+}$ |
|  | MVCT | $2Ti^{4+} + 2Ce^{3+} + Ce^{4+} + Cr^{3+} \leftrightarrow 2Ti^{3+} + Ce^{3+} + 2Ce^{4+} + Cr^{4+}$ |
| ***Pechini*** | IVCT | $Ti^{4+} + Ce^{3+} \leftrightarrow Ti^{3+} + Ce^{4+}$ |
|  |  | $3Ti^{4+} + Cr^{3+} \leftrightarrow 3Ti^{3+} + Cr^{6+}$ |

| | MVCT | $4Ti^{4+} + Ce^{3+} + Cr^{3+} \leftrightarrow 4Ti^{3+} + Ce^{4+} + Cr^{6+}$ |
|---|---|---|

The degradation of MB solution by the undoped and codoped TiO$_2$ powders as well by Degussa P25 powder, along with the relative BET surface areas, are shown in Figure 11. These data reveal the codoping is not beneficial, which is a result of the decreasing crystallinity with increasing codopant level, as shown in Figure 3. This is not an effect of surface area since Figure 6 does not correlate with the MB degradation data. However, it is likely that the low surface areas of the sol-gel powders relative to those of the Pechini powders are responsible for the differences in MB degradation since the sol-gel powders are more crystalline that the Pechini powders. In the absence of codoping, both synthesis methods produce powders of performance comparable to that of Degussa P25. More broadly, Figure 11 reveals that the sol-gel and Pechini powders exhibit two regimes of effectiveness in MB degradation.

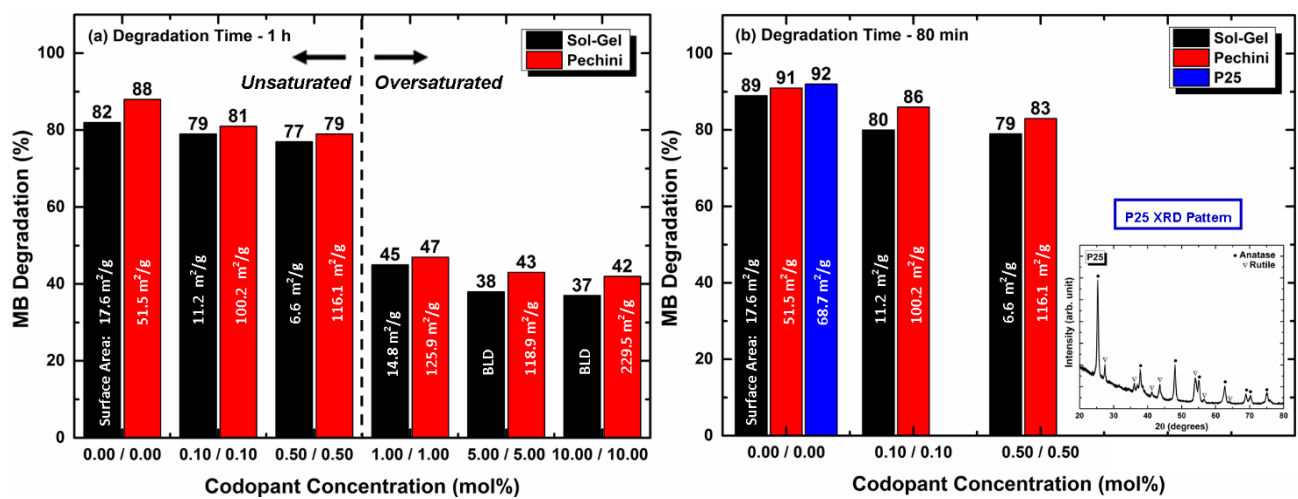

**Figure 11.** (a) Degradation of MB solutions photocatalysed by undoped and codoped TiO$_2$ powders for 1 h and (b) degradation of MB solutions by Degussa P25, undoped, and two lowest codopant concentrations for 80 min

It can be seen that the two photocatalytic regimes do not correlate precisely with any of the data considered (XRD, SEM, BET, TEM, FTIR, and IVCT). This is likely to result from the complexity introduced by the five variables (three valences, carbon content, and lattice stability) that have been mentioned and are reflected by these data. It is considered most likely that the bimodal behavior results from the solubility limits of the dopants in the TiO$_2$ lattice, where these lie between 0.50 mol%

Ce/0.50 mol% Cr and 1.00 mol% Ce/1.00 mol% Cr. When saturation effectively is achieved, this results in the onset of supersaturation (hence the progressive lattice destabilization at higher codopant levels) and the precipitation of dopant ions on the grain boundaries (which would block the photocatalytically active sites but remain BLD). Both of these would serve to decrease the photocatalytic performance. Figure 3 supports these suppositions in that peak intensity decreases can be observed in this range.

## 4. Conclusions

Ce/Cr codoped TiO$_2$ nanoparticles were synthesized using sol-gel and Pechini methods and heat treated at 400°C for 4 h. Figure 3 shows that well crystallized anatase was the only phase present in the sol-gel powders while less ordered mixed-phase anatase + rutile was present in the Pechini powders, suggesting that the latter method enhances both the solubility of Ce, which destabilizes the TiO$_2$ lattice, and hence the chemical homogeneity. The more open agglomerated morphologies, shown in Figures 4 and 5, and the higher surface areas of the Pechini derived materials, shown in Figure 6, were a reflection of the greater structural disruption from oxidative decomposition of the resin precursor, whereas the denser agglomerates from the sol-gel precursor arose from less structural disruption from pyrolysis of xerogels. The lower extent of pyrolysis of the Pechini precursor is confirmed by the FTIR data in Figure 10. The reduction effect from the pyrolysis of the carbon components was reflected in the XPS data in Figure 9, which indicated that the Ti$^{3+}$ concentration was greater in the sol-gel powders. Codoping caused a general reduction in the binding energies, which is consistent with destabilization of the lattice. The generation of Cr$^{4+}$ in the sol-gel powders and Cr$^{6+}$ in the Pechini powders from a precursor of Cr$^{3+}$ provides strong evidence for the origin of these valence changes deriving from the synergistic electron exchange involved in the possible IVCT and/or MVCT reactions shown in Tables 6 and 7. These effects, in combination with the increase in Ti$^{3+}$ concentration from codoping (and associated mobile oxygen vacancy concentration) as well as the heat generated from pyrolysis of the polyester resin of the Pechini precursor, may have promoted the anatase → rutile transformation. Since Ce is too large to allow the assumption of either substitutional or interstitial solid solubility, the concept of *integrated solubility* is introduced, whereby by the Ti site and an adjacent interstice are occupied by the large Ce ion and associated contraction of the *c* axis, as suggested in Table 4. The photocatalytic performance data in Figure 11 show that codoping was not beneficial, which is attributed largely to the effects of crystallinity (*i.e.*, lattice destabilization) and surface area. These data exhibit two regimes of mechanistic behavior, which are attributed the unsaturated solid solutions at the three lower codopant levels and supersaturated solid solutions at the three higher codopant levels. However, a full interpretation is clouded by the concurrent effects of five variables (three valences, carbon content, and lattice stability). Regardless, the present work demonstrates that the Pechini method offers a processing technique that is superior to sol-gel in that the former facilitates solid solubility and resultant chemical homogeneity.

**Acknowledgments**

The authors wish to acknowledge the financial support from the Australian Research Council (ARC) and the characterization facilities provided by the Mark Wainwright Analytical Centre at UNSW Sydney.

Wu, M.; Chen, X.; Chen, C. $WO_3$ and Ag Nanoparticle Co-Sensitized $TiO_2$ Nanowires: Preparation and the Enhancement of Photocatalytic Activity, *RSC Adv.*, **2014**, *4*, 23831-23837.